\documentclass[runningheads]{llncs}
\usepackage[T1]{fontenc}
\usepackage{graphicx}
\usepackage[T1]{fontenc} % add special characters (e.g., umlaute)
\usepackage[utf8]{inputenc} % set utf-8 as default input encoding
\usepackage{amsmath, cite, url}
\usepackage{graphicx}
\usepackage{color}
\usepackage{tabularx}
\usepackage{enumitem}
\usepackage[export]{adjustbox}
\usepackage{float}
\usepackage{lineno}
\usepackage{booktabs}
\usepackage{caption}
\usepackage{sidecap}
\begin{document}
\title{Yin-Yang: Developing Motifs With Long-Term Structure And Controllability}
\author{Keshav Bhandari\inst{1}\orcidID{0009-0008-2045-6319} \and
Geraint A. Wiggins\inst{1,2}\orcidID{0000-0002-1587-112X} \and
Simon Colton\inst{1}\orcidID{0000-0002-4887-6947}}

\authorrunning{K. Bhandari et al.}
% \institute{
% \email{k.bhandari@qmul.ac.uk}\\
% \and
% \email{geraint@ai.vub.ac.be}\\
% \and
% \email{s.colton@qmul.ac.uk}\\
% School of Electronic Engineering and Computer Science\\
% Queen Mary University of London
% % London, E1 4NS, UK
% }

% \institute{
% \email{geraint@ai.vub.ac.be}\\
% Department of Computer Science (DINF)\\
% Vrije Universiteit Brussel
% % London, E1 4NS, UK
% }

\institute{Electronic Engineering and Computer Science, Queen Mary University of London
\email{\{k.bhandari,s.colton\}@qmul.ac.uk}\\
\and
Department of Computer Science (DINF), Vrije Universiteit Brussel
\email{geraint@ai.vub.ac.be}}

\maketitle              % typeset the header of the contribution
\begin{abstract}
Transformer models have made great strides in generating symbolically represented music with local coherence. However, controlling the development of motifs in a structured way with global form remains an open research area. One of the reasons for this challenge is due to the note-by-note autoregressive generation of such models, which lack the ability to correct themselves after deviations from the motif. In addition, their structural performance on datasets with shorter durations has not been studied in the literature. In this study, we propose Yin-Yang, a framework consisting of a phrase generator, phrase refiner, and phrase selector models for the development of motifs into melodies with long-term structure and controllability. The phrase refiner is trained on a novel corruption-refinement strategy which allows it to produce melodic and rhythmic variations of an original motif at generation time, thereby rectifying deviations of the phrase generator. We also introduce a new objective evaluation metric for quantifying how smoothly the motif manifests itself within the piece. Evaluation results show that our model achieves better performance compared to state-of-the-art transformer models while having the advantage of being controllable and making the generated musical structure semi-interpretable, paving the way for musical analysis. Our code and demo page can be found at https://github.com/keshavbhandari/yinyang.

\keywords{Music Generation  \and Musical Structure \and Deep Learning.}
\end{abstract}
\section{Introduction}
When developing a short musical motif into a complete monophonic piece, composers can repeat the previous context, vary it slightly, or create something novel departing from the original motif. While composing, it is important to consider local coherence by ensuring that the next notes logically follow the previous ones, in addition to incorporating motivic variations throughout, for an overarching structure. Additionally, one may aim to produce new sections to establish global form. This is common practice in western European musical composition where people employ techniques like fragmentation, sequencing, retrograde, inversion, etc. to develop the musical idea \cite{shan2010algorithmic}. However, blindly applying these rules without considering the preceding context could be musically nonsensical. Conversely, straying too far from the original motif can lead to a lack of cohesion and unity. Finally, in some genres, repeating themes too much can be perceived as monotonous by listeners, making it undesirable \cite{burns_typology_1987}.

Computationally replicating the compositional process of developing a motif into a structured piece has been researched for decades, but remains an open challenge today \cite{bhandari2024motifs,ji_survey_2023}. Unlike storytelling with clear plot structures, the ``language'' of musical structure is more abstract and heavily relies on repeating ideas, varying them slightly, and expanding on them. This makes explicitly modelling and representing musical structure challenging.

Some studies consider the musical structure to be implicitly learnt during the training process. Approaches such as those in \cite{huang_music_2018,shih_theme_2022,zhang2021melody,de_berardinis_modelling_2020} include various enhancements to the model architecture. Data-based enhancements, where additional structural data annotations are provided to improve the model's explicit generalizability of structure, are also explored. For example, the Lookback RNN \cite{waite_e_generating_2016} encodes repeating patterns from 1 or 2 bars ago in the input to allow the RNN model to better identify repeating events in the generated melody. Similarly, the R-Transformer \cite{hu_beauty_2022} is trained on a dataset of 5 types of motif repetitions with a multiple loss objective. On the other hand, representation-based enhancements build on the idea that compact sequences would allow the model to see the complete musical work during training, enabling it to learn longer structural dependencies over the entire piece's duration \cite{hsiao_compound_2021,huang_pop_2020}. Increasing the model's context window to fit much larger sequences is yet another way to expose the complete piece to the model \cite{yu_museformer_2022,christine_musenet_2019}. Although these models show improvement over the baselines, they lack flexibility and control as their structure is automatically learnt.

Another class of models draws inspiration from the hierarchical nature of musical elements such as notes, chords, bars, phrases, and sections \cite{hornel_melonet_1997,wu_hierarchical_2018,guo_hierarchical_2021,zhang_structure-enhanced_2022}. These models use this inherent hierarchy by employing specialized networks operating at various scales to generate music. However, the window sizes across the hierarchical tiers of the neural network are fixed during training and may not match the varying hierarchy in music structure across pieces and genres.

Pre-training methods have also been explored recently with models such as MelodyGLM \cite{wu_melodyglm_2023}, which introduces a multi-task pre-training framework tailored for melodies and a fine-tuning application for melody continuation and infilling tasks. Similarly, MuPT \cite{qu2024mupt} explores the application of large language models to the pre-training of melodies encoded with a novel multi-track ABC notation. Recently, MelodyT5 \cite{wu2024melodyt5} uses the T5 transformer architecture \cite{raffel2020exploring} for pre-training with multiple objectives on a large and diverse corpus of symbolic music data. The model was evaluated on various tasks including melody generation.

More recently, a new research area within deep learning called ``sub-task decomposition'' has emerged \cite{bhandari2024motifs}. This approach breaks down the music generation process into smaller, more manageable steps (typically two). The first step aims to learn structural information from the training data, which is then used to guide the generation in the second step. Several studies have utilized this framework to generate melodies \cite{velardo_automated_2015,wu_popmnet_2020,zou_melons_2021,lu_meloform_2022,wu_compose_2023}. Among them, MeloForm \cite{lu_meloform_2022} offers user controllability but is dependent on harmony as it is as an integral part of the neural model's refinement and is therefore not suitable for monophonic melody datasets like the folk songs used in the work presented here.

While our framework consists of three distinct models -- a phrase generator, a phrase refiner, and a phrase selector, our approach differs slightly from existing sub-task decomposition works. Rather than having one stage aimed at learning structure and another at generating melodies conditioned on it, our three models are trained independently but used together during generation. Our motivation stems from the observation that transformers can generate locally coherent melodies over short time periods. However, in their pursuit of local coherence, they can sometimes make overly safe choices due to the exposure bias of the student-teacher training setup \cite{bengio2015scheduled}, leading to an unstable structure over durations longer than 2-3 minutes. Once stuck in a repetitive loop or after deviating too much from the motif, we have found that they lack the ability to correct themselves for generation of the entire piece. This issue with vanilla autoregressive transformer models may be more pronounced when they are trained on short melodies, such as those found in folk songs, but are required to generate longer sequences during inference, where they tend to struggle with maintaining structural coherence and consistent motivic development. This limitation suggests a need for approaches that allow music composers or arrangers to extend a motif into well-structured, longer sequences, an essential aspect of musical development that can often exceed the scope of a model’s training data.

We propose Yin-Yang, a transformer-based framework for developing motifs with long-term structure and controllability. Our framework employs a phrase corruption and refinement training strategy using melodic and rhythmic transformations inspired by music composition theory, producing variations of the motif. As seen in our proposed approach in Section \ref{proposed-approach}, highly similar motivic variations allow the phrase generator to adhere to the original motif, while lower similarity variations enable the creation of new sections. Our neural framework incorporates musical domain knowledge into the phrase transformation process, promoting versatility while making the generated structure semi-interpretable. Our framework offers several advantages over baseline models:
\begin{enumerate}
    \item It allows for coherent generation of melodies with long-term structure that spans much longer durations than the average length of the training data.
    \item It offers users controllability of the structure, enabling them to create new sections based on previous ones while choosing the number of phrases per section. Additionally, users can specify the type of developmental transformations, such as augmentation, fragmentation, retrograde, inversion, etc., before the actual generation, or optionally let the system choose.
    \item Our phrase selector model chooses the most suitable phrase from a pool of options, ensuring that it fits the musical context or the previous phrase. This helps maintain consistency and homogeneity within each section.
    \item As our models generate music phrase by phrase and the refiner is provided with a symbolic transformation of the previous phrase, we know exactly which phrase was transformed and how, before being refined. This allows our model to produce semi-interpretable phrases, making it easier to analyze the melodic structure. We refer to the generated structure as ``semi-interpretable'' since the neural network itself is still a black box.
\end{enumerate}

We also propose a new objective evaluation metric that measures whether a phrase was derived from the motif, indicating how smooth (cf. \cite{cambouropoulos2001pattern}) the motivic development is. We demonstrate the effectiveness of this metric alongside the Vendi score \cite{friedman2023vendi} from the machine learning literature, which can quantify the diversity of the generated outputs.
\vspace{-0.2cm}

\section{Proposed Approach}
\label{proposed-approach}
At the core of our algorithm are the phrase generator and refiner models. The phrase generator is analogous to an autoregressive language model, which in the musical sense learns to generate content in the style of the dataset on which it is trained. As the phrase generator struggles with repetitions and novel motifs, the phrase refiner assists it structurally in either smoothly bringing back a variation of the motif or in defining a new motif for the next section of the piece. The phrase refiner is trained on a self-supervised corruption refinement strategy. During generation, transformations of the phrase containing the motif are presented to the phrase refiner to generate a meaningful variation with respect to the context.
% \vspace{-0.5cm}

\subsection{Phrase Generator}
% \vspace{-0.2cm}
Let the set of musical phrases\footnote{https://musictheory.pugetsound.edu/mt21c/PhraseSection.html} in a dataset be $\{P_1, P_2, \ldots, P_N\}$. The phrase generator $h_\theta$ is an encoder-decoder transformer \cite{vaswani2017attention} model that is provided previous phrases as input to produce the next phrase as follows:
\vspace{-0.5\baselineskip}
\begin{equation}
\begin{split}
P_{N+1} &= h_\theta(P_k, P_{k+1}, \ldots, P_N, \quad C) \\
&\text{where } k < N, \text{ and } C \text{ are conditional tokens} 
% j \in \{1, 2, \dots, m\} 
\label{eq:generator}
\end{split}
\end{equation}
% \vspace{-0.7\baselineskip}

We condition the model on the type of corruption, key and time signature, phrase length and the cadence of the target phrase $P_{N+1}$ to make it controllable at generation time. These conditional tokens are concatenated in the sequence.

% \vspace{-0.2cm}
\subsection{Phrase Refiner}
% \vspace{-0.2cm}
During training, we take two consecutive phrases from our dataset (e.g. $P_i$ and $P_{i+1}$) and corrupt $P_{i+1}$ with a corruption function $g_j$ (see Section \ref{transformation}) chosen at random from the set $\{g_1, g_2, \ldots, g_J\}$ such that:
% \vspace{-0.5\baselineskip}
\begin{equation}
    P_{\text{corrupted}} = g_j(P_{i+1}).
\end{equation}
% \vspace{-0.7\baselineskip}

Here, $P_{\text{corrupted}}$ represents the corrupted phrase $P_{i+1}$, which may have lost some of its properties of being called a musical phrase due to the applied corruptions. The phrase refiner model $r_\theta$ consists of an encoder-decoder transformer architecture that takes the unchanged $P_{i}$ along with $P_{\text{corrupted}}$ and conditional tokens $C$ with a training objective to reproduce $P_{i+1}$, the clean version of the second phrase, denoted as:
% \vspace{-0.5\baselineskip}
\begin{equation}
    P_{i+1} = P_{\text{refined}} = r_\theta(P_{i}, \quad P_{\text{corrupted}}, \quad C).
\end{equation}
\vspace{-0.7\baselineskip}

During generation, suppose that we have a given phrase $P_1$ containing the motif and a set of phrases generated from the phrase generator $\{P_2, \ldots, P_{N}\}$. Then, to produce a variation to generate the phrase number $P_{N+1}$, we select $P_1$ and transform it with a transformation function $f_j$ (see Section \ref{transformation}) selected at random or by the user from the set $\{f_1, f_2, \ldots, f_J\}$ such that: 
% \vspace{-0.5\baselineskip}
\begin{equation}
    P_{\text{transformed}} = f_j(P_1).
\end{equation}
\vspace{-0.7\baselineskip}

The difference between a corruption and a transformation function lies in the time of its usage; corruptions are applied during training, whereas transformations are applied during generation. It is important to note that the chosen transformation function $f_j$  should be logically commensurate with the corruption function $g_j$ that the phrase refiner model $r_\theta$ saw during training. Section \ref{transformation} explains this. After the transformation, we pass the last generated phrase $P_N$ along with the transformed phrase $P_{\text{transformed}}$ and user-defined conditional tokens $C$ to $r_\theta$ to produce a refined variation of the motif $P_{N+1}$, denoted as:
% \vspace{-0.5\baselineskip}
\begin{equation}
    P_{N+1} = P_{\text{refined}} = r_\theta(P_N, \quad P_{\text{transformed}}, \quad C). \label{eq:refiner}
\end{equation}
\vspace{-2.0\baselineskip}

\begin{figure}
 \centerline{
 \includegraphics[width=1.0\columnwidth, height=3.0in, left]{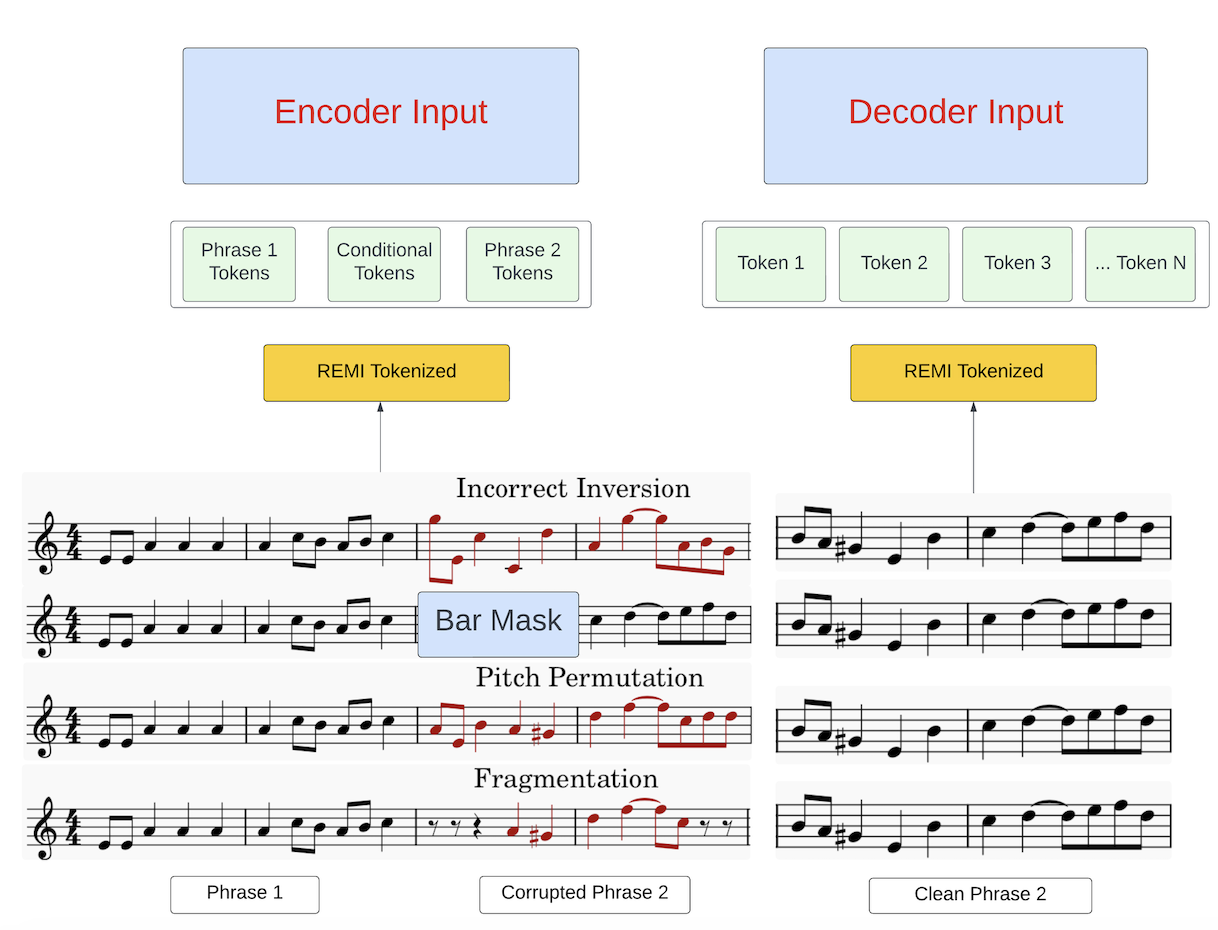}}
 \captionsetup{justification=raggedright, font=small}
 \caption{The phrase refiner encoder takes in phrase 1 along with the conditional tokens and the corrupted phrase 2 to generate the clean phrase 2 with its decoder. The conditional tokens include the type of corruption, key and time signatures, along with phrase length and cadence of the phrase.}
 \vspace{-1.0\baselineskip}
 \label{fig:corruption-refinement-training}
\end{figure}

% \vspace{-0.2cm}
\subsection{Transformation and Corruption Functions} \label{transformation}
% \vspace{-0.2cm}
During training, we corrupt every following phrase shown to the phrase refiner model based on the corruption functions shown below. We provide the model with corruption tokens to indicate the type of corruption. This can be seen visually in Figure \ref{fig:corruption-refinement-training}. These corruption tokens are then paired with the transformations at generation time to refine the transformed phrases to better match the preceding context. 
\begin{itemize}
    \item \textbf{Fragmentation}\footnote{https://musictheory.pugetsound.edu/mt21c/FragmentSection.html}: Returns a random crop or bar from the phrase.
    \item \textbf{Permutation}\footnote{https://en.wikipedia.org/wiki/Permutation\_(music)}: Randomly shuffles either note pitch, duration, or both.
    \item \textbf{Incorrect Inversion}: Changes the pitch value of notes in the phrase randomly by [-4,4] semitones.
    \item \textbf{Melodic Stripping \& Addition}: Melodic stripping omits out a note at random with a 50\% probability whereas melodic addition introduces a new note from the same key signature of the phrase with a 50\% probability.
    \item \textbf{Masking}: Masks either the pitch, duration (and note's position) or an entire bar from the phrase.
    \item \textbf{Same Note Modification}: Either removes or adds a note of the same pitch value to the phrase.
\end{itemize}

Inspired by melodic and rhythmic development theories in Western European music composition \cite{marvin1989generalized}, we provide several transformations to a motif during generation time to develop the piece. 
The phrase refiner takes in the clean phrase along with the transformed one and the corruption token which indicates the direction in which to refine the transformation.

\begin{itemize}
    \item \textbf{Fragmentation}: Similar to the fragmentation corruption - notes outside the fragment are omitted.
    \item \textbf{Permutation}: Similar to permutation corruption.
    \item \textbf{Inversion}\footnote{\label{footnote1}https://musictheory.pugetsound.edu/mt21c/MelodicAlteration.html}: Changes the pitch contour by inverting intervals either with chromatic or modal inversion.
    \item \textbf{Retrograde}\footref{footnote1}: Reverses in time either note pitch or both note pitch and duration.
    \item \textbf{Augmentation \& Diminution}\footref{footnote1}: Augmentation doubles the duration of notes. Diminution halves the duration of notes.
    \item \textbf{Reduction}: Omits note if next note has same pitch value.
    \item \textbf{Masking}: Similar to masking corruption. The goal is to produce a variation with the masked notes, while keeping the other notes intact.
\end{itemize}

% \vspace{-0.5cm}

\begin{table}
 \begin{center}
 \small
 \begin{tabular}{|p{5.5cm}|p{6.5cm}|}
  \hline
  \textbf{High Similarity Transformations} & \textbf{Logical Corruptions} \\
  \hline
  Fragmentation & Fragmentation \\
  \hline
  Permute Duration & Permute Duration \\
  \hline
  Augmentation & Melodic Stripping \\
  \hline
  Diminution & Melodic Addition; Same Note Modification \\
  \hline
  Reduction & Melodic Addition; Same Note Modification \\
  \hline
  Duration Masking & Duration Masking \\
  \hline
  Bar Masking & Bar Masking \\
  \hline
 \end{tabular}
\end{center}
% \vspace{-0.8\baselineskip}
\captionsetup{justification=raggedright, font=small}
 \caption{High similarity transformations \& corruptions.}
 \vspace{-3.0\baselineskip}
 \label{tab:Table 1}
\end{table}

\begin{table}
 \begin{center}
 \small
 \begin{tabular}{|p{5.5cm}|p{6.5cm}|}
  \hline
  \textbf{Low Similarity Transformations} & \textbf{Logical Corruptions} \\
  \hline
  Permute Pitch  & Permute Pitch \\
  \hline
  Permute Pitch \& Duration & Permute Pitch \& Duration \\
  \hline
  Real Inversion & Incorrect Inversion \\
  \hline
  Tonal Inversion & Incorrect Inversion \\
  \hline
  Retrograde Pitch & Incorrect Inversion \\
  \hline
  Retrograde Pitch \& Duration & Incorrect Inversion; Permute Pitch \& Dur. \\
  \hline
  Pitch Masking  & Pitch Masking \\
  \hline
 \end{tabular}
\end{center}
% \vspace{-0.8\baselineskip}
\captionsetup{justification=raggedright, font=small}
 \caption{Low similarity transformations \& corruptions.}
 \label{tab:Table 2}
 \vspace{-2.0\baselineskip}
\end{table}

\begin{figure}
\centering
% \hspace{-1cm}
 \includegraphics[width=1.0\columnwidth, height=2.7in, left]{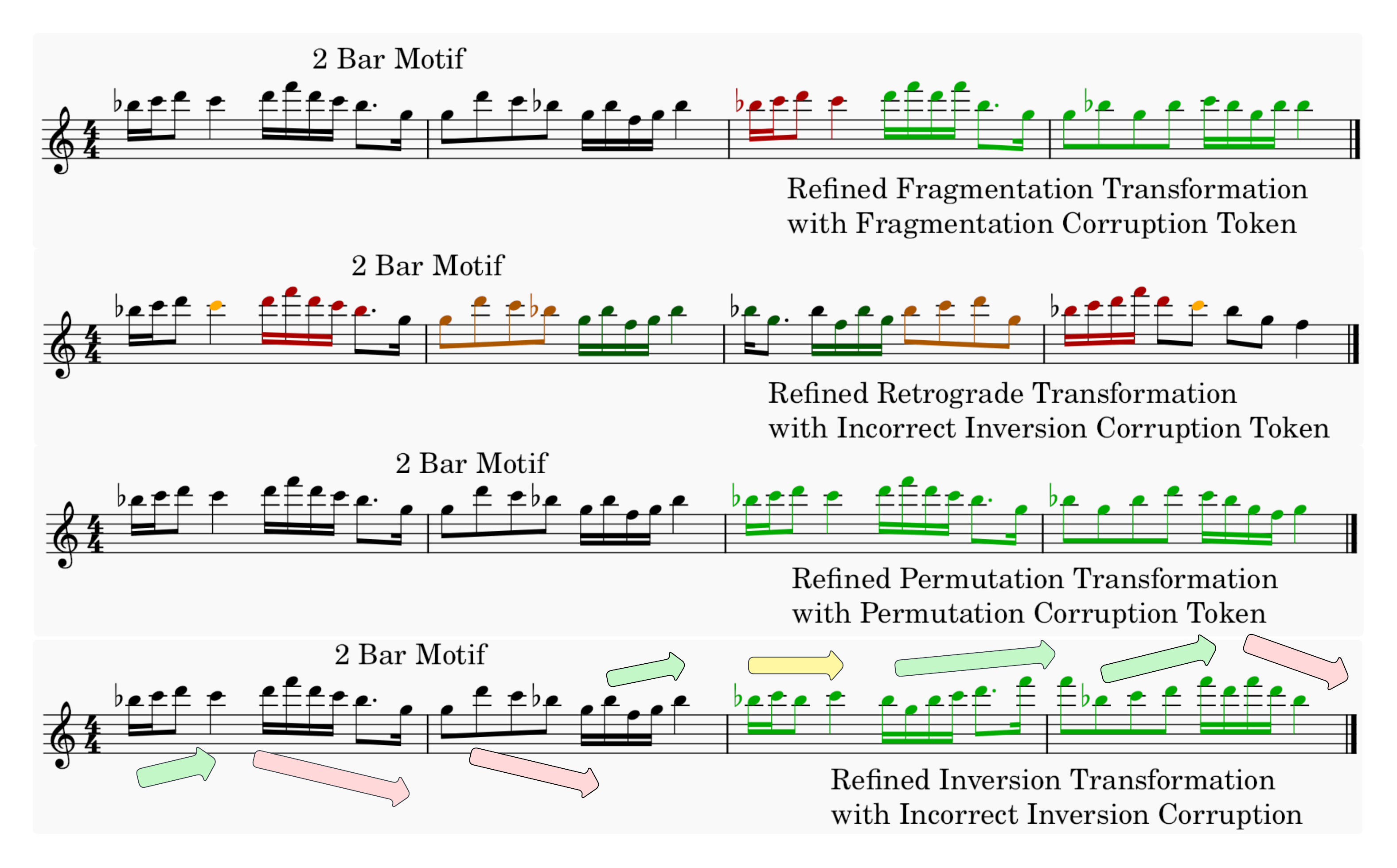}
 % \vspace{-0.5cm}
 \captionsetup{justification=raggedright, font=small}
 \caption{Semi-interpretable examples of refined transformations of a given 2 bar motif: Fragmentation retains the red fragment while the green notes vary. For retrograde, an incorrect inversion corruption produces a retrograded motivic variation. In inversion, the generated pitch contour inversely relates to the motif in a non-strict way while adhering to prior context as seen with the yellow arrow.}
 \vspace{-0.4cm}
 \label{fig:interpretable}
\end{figure}

Tables \ref{tab:Table 1} and \ref{tab:Table 2} highlight the transformations by their similarities to the original motif and their corresponding corruption tokens that are shown to the refiner model at generation time. Most transformations at generation time are similar to corruptions during training. We match the transformations with corruptions that make logical sense, while also sounding pleasant in our experiments. For example, as seen in Table \ref{tab:Table 1}, augmentation increases the duration of each note and is complemented by the melodic stripping corruption token, which helps the model add new notes in between to create an interesting variation. Similarly, we find that retrograde works well with incorrect inversion corruption tokens as the pitch contour is reversed, but not strictly, so that it better matches the preceding context. More examples of these are shown in Figure \ref{fig:interpretable}.

During training, we use fragmentation corruption with a 20\% probability and combine it with the other non-masking corruptions. Masking corruptions are shown with a 20\% probability at training time but are not combined with other corruptions. For multiple corruptions that are shown in the tables above, only one is selected at random to be shown to the phrase refiner model.

\subsection{Phrase Selector}\label{sec:phrase-selector}
% \vspace{-0.2cm}
The objective of the phrase selector model is to choose the most suitable phrase from a pool of generations for a given musical context within a section. One way to define ``most suitable'' may be a phrase that is the most likely generation of a previous phrase. $K$ different $P_{n+1}$ phrases are generated from equations \ref{eq:generator} and \ref{eq:refiner} above by varying the models' temperature parameter. Let $f_s(P_{n+1})$ represent the function that scores each of these generated phrases. The phrase with the highest score, denoted $\hat{P}_{n+1}$, is selected as the most likely phrase given the context. Therefore, we have the following:
% \vspace{-0.5\baselineskip}
\begin{equation}
\hat{P}_{n+1} = \arg\max_{P} f_s(P_{n+1})
\end{equation}
\vspace{-0.8\baselineskip}

To train the phrase selector model, we take two consecutive phrases from the same song and treat them as positive labels, while taking another pair from different songs as negative labels. We use a transformer encoder-only model with a binary classification objective for our setup. Note that while the phrase selector is put in place to ensure consistency of phrase-based generations, the objective function can also include other metrics based on musical tension, similarity, etc.

\subsection{The Yin-Yang Framework}\label{sec:framework}
% \vspace{-0.2cm}

For a particular musical section \cite{leichtentritt1951musical}, the Yin-Yang framework operates by first calling the phrase generator to generate an initial set of $N$ phrases based on the given prompt. The phrase refiner is then invoked, taking the original motif (first phrase) and applying a \textit{high} similarity transformation to generate a similar output. This transformed motif is then refined by the phrase refiner with respect to the last phrase generated from the phrase generator, producing the ($N+1$)th phrase. The ratio of phrases generated by the phrase generator versus those refined by the phrase refiner is controlled by a user-defined parameter G:R applicable within a particular section. For subjective experiments, this ratio was set to 2:1, meaning that the phrase generator produced 2 phrases before the refiner generated the 3rd phrase through motivic transformation and refinement. For the 4 phrases in each section, the phrases in this ratio were generated as [M,G,G,R,G], where M, G and R are the motif, generator and refiner models, respectively. The phrase selector is used on both models to select the most suitable generated phrase from a pool of 5 phrases ($K=5$) with varying temperatures.

\begin{figure}
 \centerline{
 % \hspace{-1cm}
 \includegraphics[width=1.0\columnwidth, height=2.0in, left]{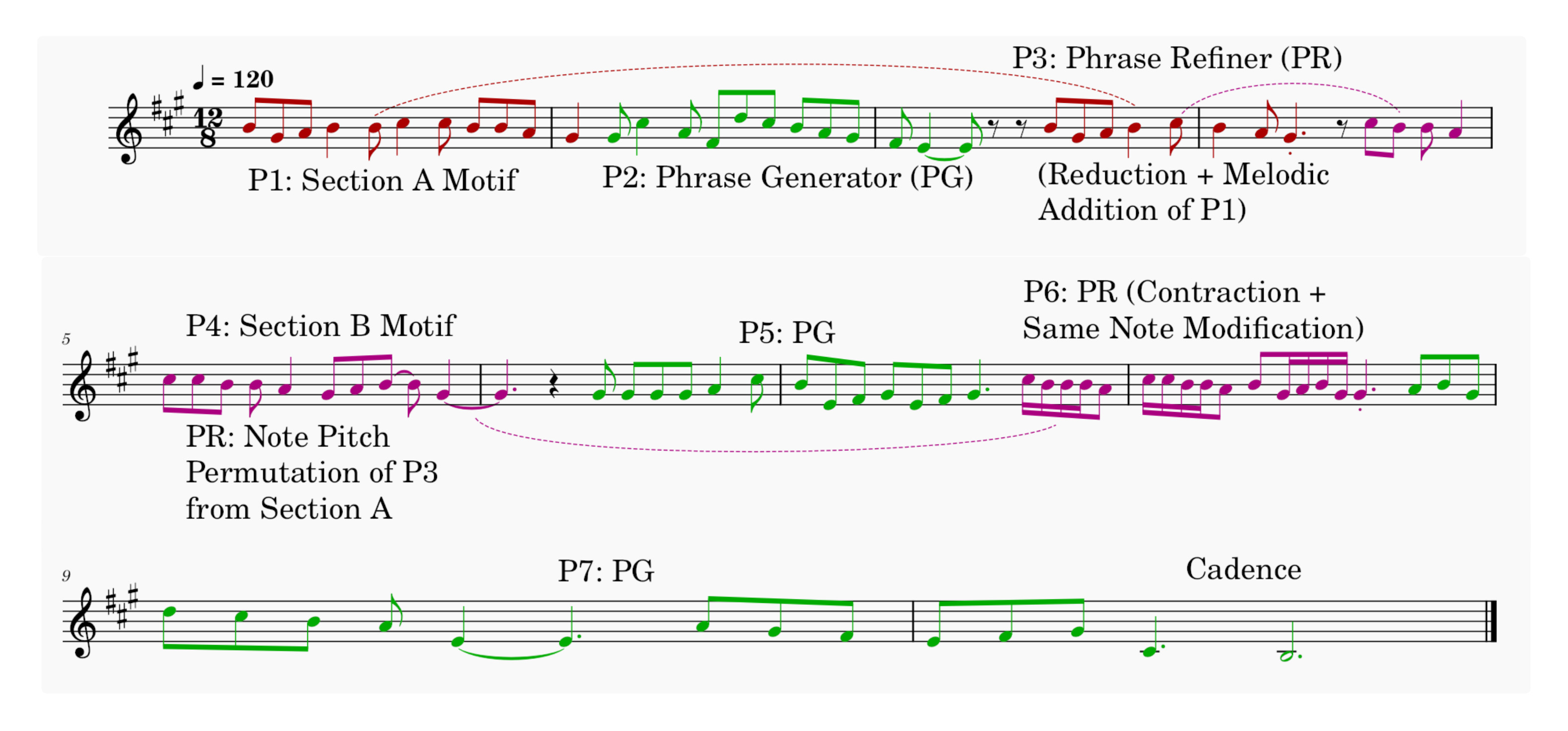}}
 \captionsetup{justification=raggedright, font=small}
 \caption{Semi-interpretable generation framework of Yin-Yang with 2 sections AB consisting of 3 and 4 phrases respectively.}
 \vspace{-0.5cm}
 \label{fig:generation-framework}
\end{figure}

To generate the same section (e.g., AA), we copy the motif again and randomly sequence it an octave higher or lower, or keep it the same. The phrase generator is only allowed to see phrases as context within the particular section it generates for. To produce new sections, we take any arbitrary phrase from the previous sections and apply a \textit{low} similarity transformation and refine it with respect to the last phrase of the previous section. To add more novelty to the new motif for the new section, we can limit the number of notes of the last phrase that the refiner sees as context, to reduce its dependence upon it. In our case, we limit it to only the last bar. The length of the new motif of the new section is conditioned to be between 9-16 notes. We also have the flexibility to modulate to new key signatures as both models condition on this information. A small example of the framework comprising 2 sections with 7 total phrases (A3B4) and a PG ratio of 1:1 can be seen in Figure \ref{fig:generation-framework}. For our user study, we generate 20 phrases in ABACA form. Our aim is to have homogeneous or highly similar phrases within a section but heterogeneous phrases between new sections.

% \vspace{-0.5cm}
\section{Experimental Configurations}\label{experiments}

\subsection{Dataset}\label{dataset} 
% \vspace{-0.2cm}
We used three monophonic folk song datasets in our study:
MTC-FS-INST 2.0 \cite{van2019meertens}, MTC-ANN-2.0.1 \cite{van2016meertens} and ESSEN Folksong Collections\footnote{http://www.esac-data.org/}. All three datasets have phrase annotations in JSON files which can be obtained here\footnote{https://zenodo.org/records/3551003} along with their documentation\footnote{https://pvankranenburg.github.io/MTCFeatures/introduction.html}. We preprocessed the files by extracting each note's pitch and duration information, along with the key and time signatures. We included all key and time signatures in this study, but excluded songs with only a single phrase. Table \ref{tab:Table3} shows the statistics of all three datasets after preprocessing. As seen in the last two columns, the average song duration is quite short, typical of folk songs. Although some songs in the dataset develop the initial musical idea, this development is typically less extensive than in other genres, such as classical music, reflecting a characteristic feature of folk songs rather than a drawback. We chose this dataset because it highlights the effectiveness of our proposed methods in elaborating on a musical idea, even when fully developed songs are limited. To train all the models, we split the data into training, validation and test sets which account for 23,733 (\textasciitilde 90\%), 2,649 (\textasciitilde 10\%) and 100 files respectively. Each set comprises unique phrases  with no overlap between them.

\begin{table}
\centering
\small
\begin{tabular}{p{1.8cm}>{\centering\arraybackslash}p{0.8cm}>{\centering\arraybackslash}p{3.5cm}>{\centering\arraybackslash}p{3.5cm}}
\hline
Dataset & \#Songs & Avg. Phrases per Song & Avg. Notes per Song \\
\hline
\hline
MTC-FS & 360 & 6.19 & 71.74 \\
\hline
MTC-ANN & 18,109 & 5.25 & 48.07 \\
\hline
ESSEN & 8,469 & 5.61 & 52.61 \\
\hline
Combined & 26,938 & 6.0 & 65.40 \\
\hline
\end{tabular}
\vspace{0.5\baselineskip}
\captionsetup{justification=raggedright, font=small}
\caption{Statistics of all datasets after excluding single phrases.}
\label{tab:Table3}
\vspace{-1.5\baselineskip}
\end{table}

\vspace{-0.5cm}
\subsection{Model Implementation Details} \label{model-implementation}
% \vspace{-0.2cm}
The phrase generator \textbf{(PG)} and phrase refiner \textbf{(PR)} models have a 4-layered encoder-decoder transformer architecture with 4 attention heads, hidden size of 512, and intermediate size of 2048. PG has a longer encoder sequence length of 2048 tokens to accommodate longer input contexts compared to 512 tokens for the PR as it only looks back at the last phrase. Both decoders have a context window of 512 as they generate a single phrase at a time. The phrase selector \textbf{(PS)} transformer is an encoder-only architecture with the same configurations as the encoder of the PR model but with a context window of 1024 tokens. All three models are trained for 30 epochs (or until early stopping) using the Adam optimizer with a learning rate of 0.0001, weight decay of 0.01, and maximum gradient norm of 3.0. We use the REMI sequence encoding \cite{huang_pop_2020} and pitch shifting (between $\pm 12$ semitones) data augmentation for all three models.

% \vspace{-0.5cm}
\subsection{Baselines And Ablations}
% \vspace{-0.2cm}
To demonstrate our framework's effectiveness against state-of-the-art transformer-based models that implicitly learn musical structure, we select two baselines. The first is the Music Transformer (\textbf{MT}) \cite{huang_music_2018}, which incorporates \textit{relative positional attention}, a model-based enhancement. The second is the Compound Word Transformer (\textbf{CP}) \cite{hsiao_compound_2021}, aimed at compact sequence representations, a feature-based enhancement. We train both models from scratch on the datasets used in this study with configurations similar to the PG model described above. Our goal is to demonstrate that, in comparison to the baselines, the Yin-Yang framework can extend motifs into longer, well-structured melodies without the need for explicit training on extended sequences.

Additionally, we aim to investigate the impact of the PS model on our framework's performance. To this end, we evaluate an ablated version of our framework, both with and without the PS model. We denote the complete Yin-Yang model with the selector as \textbf{YY}, and the ablated version without as \textbf{YYA} in all the experiments below. 
We re-emphasize the goal of this study is to go beyond the real dataset which has limited motivic development and new sections for the majority of the pieces. To this end, comparing our framework against the test set is not meaningful for the goal of our study.
% \vspace{-0.2cm}

\section{Experimental Evaluation}
% \vspace{-0.1cm}
\subsection{Objective Evaluation}

Previous studies \cite{guo_hierarchical_2021} on structure generation used the COSIATEC compression ratio \cite{meredith_cosiatec_2013} and self-similarity matrices \cite{wu_compose_2023} to discover repeated segments, while \cite{wang2023motif} counted motif repetitions throughout the length of the generated piece. However, a major limitation is that these metrics can be inflated by excessive repetition of a pattern. For example, simply looping a phrase would yield high scores despite a lack of true structural development. Additionally, these metrics cannot quantify the gradual transformation and progression of the motif throughout the piece, a crucial aspect of motivic development in composition.

Musical development requires manifesting the motif throughout a section. We design an objective metric to evaluate the smooth evolution from the initial motif and the extent to which the generated material is derived from the prompt. To this end, we trained a structural derivation (\textbf{SD}) model with a similar architecture to the phrase selector model but with a different training methodology. Instead of considering two consecutive phrases from the same song as a positive example, we allow the model to treat any pair of phrases within the same song as positive, while the negative examples remain phrases from different songs. During the evaluation, given a set of generated phrases $\{P_1, P_2, \ldots, P_N\}$, where $P1$ is the motif, we consider the average of the SD model's probabilities over phrase pairs $\{P1, P2\},\{P1, P3\}, \ldots, \{P1, P_N\}$.

A moderate to high SD score indicates a smooth evolution of the generated music while adhering to the initial motif, quantifying its manifestation throughout the piece. In contrast, a low SD score may signify undesirable deviations from the motif. However, excessive repetition of a phrase can artificially inflate the SD score. Therefore, we complement the SD score with the Vendi score \cite{friedman2023vendi}, a recently proposed general diversity evaluation metric for machine learning outputs. To extract the Vendi score, we analyze the similarity matrix computed from the embeddings of the last hidden state output of the SD model in $\{P1, P2\},\{P1, P3\}, \ldots, \{P1, P_N\}$. Passing this matrix through the Vendi score function gives us a diversity score for all the phrases in the song. A high Vendi score indicates highly diverse outputs, while a low score suggests over-repetition. Note that in preliminary experiments with the SD model, we were satisfied with its accuracy in detecting a smooth transition of motifs. Furthermore, as described below, we validated its usage through a listening study, investigating melodic smoothness with respect to the motif and unwanted repetitions. 

We stress the importance of viewing both these scores simultaneously for a more holistic evaluation of the model's performance. As there is no easy way to differentiate between musical sections with the MT and CP baselines, we generate only Section A, comprising 20 phrases in total with the YY and YYA models in the objective study for a fair comparison. We also provide the pitch range and unique pitch averages per phrase as objective metrics to indicate the pitch contour height and pitch diversity of generations. The results are given in Table \ref{tab:objective-study-results} with the analysis provided in Section \ref{analysis}.
\vspace{-0.5\baselineskip}

\begin{table}
\centering
% \small
\setlength{\tabcolsep}{12pt}
\begin{tabular}{|c|c|c|c|c|}
\hline
 & $SD$ $\uparrow$ & $V$ $\uparrow$ & $PR$ & $UP$ \\
\hline
CP & 0.48 & 2.02 & 7.79 & 4.56 \\
MT & 0.67 & 1.54 & 9.39 & 5.42 \\
YY & 0.63 & 1.79 & 9.32 & 5.79 \\
YYA & 0.62 & 1.87 & 9.72 & 5.68 \\
\hline
\end{tabular}
\vspace{0.5\baselineskip}
\captionsetup{justification=centerlast}
\caption{Objective study results. \\
% \vspace{0.5\baselineskip}
(\textbf{S}tructure \textbf{D}erivation, \textbf{V}endi, \textbf{P}itch \textbf{R}ange, \textbf{U}nique \textbf{P}itches)}
\label{tab:objective-study-results}
\vspace{-3.5\baselineskip}
\end{table}

\subsection{Subjective Evaluation} \label{subsec: subjective}
% \vspace{-0.2cm}
We invited 22 people to participate in our listening test. Among these, 8 participants had at least some level of musical knowledge. We randomly chose 4 musical samples from each of the 4 models (CP, MT, YY and YYA). We conducted a mean opinion score (MOS) test in which participants were asked to provide their ratings from 1 (bad) to 5 (good) on the following aspects:

\noindent\textbf{Coherence ($Ch$)}: How well does the melody follow the initial prompt, and unfold smoothly throughout the piece? \\
\textbf{Similarity ($Sm$)}: Does the melody have similarities to the initial prompt as it progresses? \\
\textbf{Structureness ($S_a$)}: Does the melody have recurring motifs, phrases, and sections and reasonable musical development? \\
\textbf{Structureness ($S_b$)}: Are there multiple distinct musical ideas present within the melody? \\
\textbf{Expectedness ($E$)}: Are there unwanted repetitions of phrases in the melody? \\
\textbf{Richness ($R$)}: Is the melody intriguing and diverse? \\
\textbf{Overall ($O$)}: Subjectively, how much do you like the music?

% Table spanning two columns
\begin{table}
% \small
% \hspace{-0.5cm}
\setlength{\tabcolsep}{12pt}
\begin{tabular}{@{}|c|c|c|c|c|c|c|c|@{}}
    \hline
     & $Ch$ $\uparrow$ & $Sm$ $\uparrow$ & $S_a$ $\uparrow$ & $S_b$ $\uparrow$ & $E$ $\uparrow$ & $R$ $\uparrow$ \\
    \hline
    CP & 2.74 & 2.80 & 2.88 & 2.83 & 2.82 & 2.83 \\
    MT & 2.73 & 2.68 & 2.71 & 2.47 & 2.27 & 2.45 \\
    YY & \textbf{3.52} & \textbf{3.58} & \textbf{3.59} & \textbf{3.39} & 3.29 &\textbf{ 3.41} \\
    YYA & 3.36 & 3.49 & 3.58 & 3.36 & \textbf{3.32} & 3.30 \\
    \hline
\end{tabular}
\vspace{0.5\baselineskip}
\captionsetup{justification=centerlast}
\caption{User study mean opinion score results. \\(\textbf{C}o\textbf{h}erence, \textbf{S}i\textbf{m}ilarity, \textbf{S}tructureness, \textbf{E}xpectedness, \textbf{R}ichness)}
 \vspace{-1.5cm}
\label{table:user-study-results}
\end{table}

\begin{figure}
 \centering
 \includegraphics[width=0.8\columnwidth, height=2.0in, left]{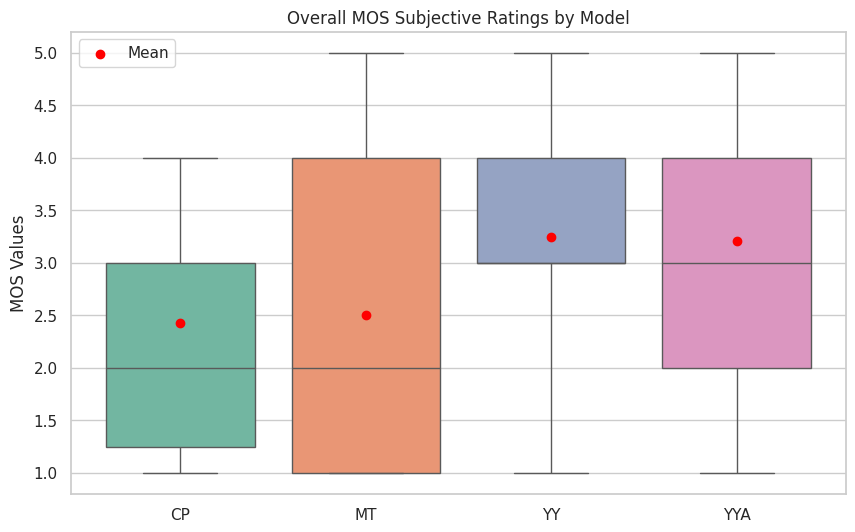}
 \captionsetup{justification=raggedright}
 \caption{Boxplot of overall MOS ratings by model group}
 \vspace{-1.0cm}
 \label{fig:mos-ratings}
\end{figure}

\subsection{Analysis and Discussion}
\label{analysis}
% \vspace{-0.2cm}
We conducted a one-tailed ANOVA test across all 4 models for each metric in Table \ref{table:user-study-results}. The results were statistically significant at $\alpha$ = 0.001, indicating differences in means between models for each metric. Pairwise comparisons using the Tukey HSD test between all models revealed significantly better results for the YY and YYA models compared to the baselines. However, pairwise comparisons between YY and YYA did not show significant differences.

The low SD score of the CP baseline in Table \ref{tab:objective-study-results} suggests an inability to manifest the motif throughout, often lacking motif-centered structural development. Its high Vendi score indicates that generations diverge too much from the initial prompt, supported by the low coherence ($Ch$) and similarity ($Sm$) participant scores in Table \ref{table:user-study-results}. The MT baseline in contrast tends to produce overly safe, motif-following notes at first glance, as seen in the higher SD score. However, excessive repetition throughout leads to low participant scores in structural metrics, confirmed by poor expectedness ($E$) and Vendi scores that indicate a lack of diversity. Participants found MT overly repetitive compared to CP, with significantly lower expectedness scores in pairwise tests.

YY and YYA achieve significantly better results in all subjective evaluation metrics and a statistically higher overall MOS score as seen in Figure \ref{fig:mos-ratings}. The moderately high SD score along with the Vendi scores further suggest motif-centric structural development with varied outputs. While there is no theoretical upper bound on the Vendi score, we can interpret CP's score of 2.02 to indicate outputs that are so varied so as not to adhere to the motif and MT's score of 1.54 to be too repetitive, which are both confirmed by our user studies. Thus, YY and YYA models are placed appropriately in the middle for both metrics, making them neither overly repetitive nor structurally lacking. 

Although the YY model performs slightly better than the ablated YYA for several subjective questions, the differences were not statistically significant. Objectively, YY's slightly higher SD score complemented by the lower Vendi score align with the goal of the phrase selector choosing the most suitable next phrase, making the generated section more homogeneous. Removing the phrase selector in YYA gives more freedom to the PG and PR models to generate without an intermediary selection process. Thus, users can control the degree of homogeneity within the section by keeping the PS model in. 
% \vspace{-0.2cm}

\section{Conclusions and Future Work}
% \vspace{-0.2cm}
In this work, we proposed Yin-Yang, a framework for developing a musical motif into a structured piece with controllability and long-term coherence. Our approach employs three models: a phrase generator to produce initial melodic phrases, a phrase refiner that applies transformations inspired by Western European music theory to produce motivic variations that logically develop the motif, and a phrase selector that chooses the most likely phrase sequence. Through objective and subjective evaluations, we demonstrated that our framework generates melodies with smooth structural evolution from the initial motif while maintaining diversity, outperforming state-of-the-art transformer models. The semi-interpretable generative nature of our approach also allows for musical analysis of the generated pieces. Yin-Yang takes a step towards computationally emulating the creative process of motivic development in musical composition. 

In future work, we plan to use the Yin-Yang framework on complex datasets, including polyphonic music. The refiner and generator models could be conditioned on pitch contour, musical tension, chord progressions, etc. to give the music more direction. We can also equip the phrase selector with additional constraints in its search for the most suitable generation. Integrating these features could bring us closer to developing an intelligent music generation system capable of considering multiple decision points to enhance musical output. 
% We also plan to experiment with the generation of longer variations of musical themes and apply the corruption refinement method to genre style transfer to convert the style of a piece while adhering to the original musical structure.

\section*{Acknowledgements}
This work was supported by UKRI and EPSRC under grant EP/S022694/1. 
We express our thanks to our reviewers for their insightful critiques, which greatly strengthened the work.

\clearpage

\bibliographystyle{splncs04}
\bibliography{references}

\end{document}